\documentclass[pra,twocolumn,showpacs,preprintnumbers,amsmath,amssymb,superscriptaddress]{revtex4-1}

\usepackage{graphicx}
\usepackage{dcolumn}
\usepackage{bm}
\usepackage{mathbbol}
\usepackage{color}
\usepackage[FIGTOPCAP,raggedright,nooneline]{subfigure}

\newcommand{\Bra}{\langle}
\newcommand{\Ket}{\rangle}

\newcommand{\ignore}[1]{}

\newcommand{\cnb}[1]{{\color{black} #1}}
\newcommand{\cnbb}[1]{{\color{black} #1}}

\begin{document}


\title{Degenerate optomechanical parametric oscillators:
\\
cooling in the vicinity of a critical point}

\author{Peter Degenfeld-Schonburg}
\email{peter.degenfeld-schonburg@ph.tum.de}
\affiliation{Technische Universit{\"a}t M{\"u}nchen, Physik Department, James Franck Str., 85748 Garching, Germany}
\author{Mehdi Abdi}
\affiliation{Technische Universit{\"a}t M{\"u}nchen, Physik Department, James Franck Str., 85748 Garching, Germany}
\author{Michael J. Hartmann}
\affiliation{Institute of Photonics and Quantum Sciences, Heriot-Watt University, Edinburgh, EH14 4AS, United Kingdom}
\author{Carlos Navarrete-Benlloch}
\email{carlos.navarrete@mpq.mpg.de}
\affiliation{Max-Planck-Institut f\"ur Quantenoptik, Hans-Kopfermann-str. 1, 85748 Garching, Germany}

\date{\today}

\begin{abstract}
Degenerate optomechanical parametric oscillators are optical resonators in which a mechanical degree of freedom is coupled to a cavity mode that is nonlinearly amplified via parametric down-conversion of an external pumping laser. Below a critical pumping power the down-converted field is purely quantum-mechanical, making the theoretical description of such systems very challenging. Here we introduce a theoretical approach that is capable of describing this regime, even at the critical point itself. We find that the down-converted field can induce significant mechanical cooling and identify the process responsible of this as a \cnb{cooling-by-heating} mechanism. Moreover, we show that, contrary to naive expectations and semi-classical predictions, cooling is not optimal at the critical point, where the photon number is largest. Our approach opens the possibility for analyzing further hybrid dissipative quantum systems in the vicinity of critical points.
\end{abstract}

\maketitle

%
%
\section{Introduction}

Degenerate optical parametric oscillators (DOPOs) consist of a driven optical cavity containing a crystal with second-order optical nonlinearity \cite{Boyd03,BlueBook,CarmichaelBook2,NavarretePhDthesis}. Down-conversion in the crystal can generate a field at half the frequency of the driving laser and classical electrodynamics predicts that such field will start oscillating inside the cavity only if the external laser power exceeds some threshold value, where the nonlinear gain can compensate for the cavity losses. A fully quantum-mechanical theory, on the other hand, reveals that even below threshold the down-converted field is not vacuum, but a squeezed field whose quantum correlations increase as the threshold is approached.

Recent developments in the fabrication of crystalline whispering gallery mode (cWGM) resonators \cite{Ilchenko03,Ilchenko04,Savchenkov07,Furst10,Furst10b,Hofer10,Furst11,Beckmann11,Werner12,Fortsch13,Marquardt13,Fortsch14,Fortsch14b} have opened the way to study the intracavity interplay between down-conversion and optomechanics \cite{Aspelmeyer13}, a setup that we will refer to as \textit{degenerate optomechanical parametric oscillator} (DOMPO). So far, it has been shown that the presence of down-conversion in an optomechanical cavity can help enhancing mechanical cooling \cite{Agarwal09}, normal mode splitting \cite{Agarwal09bis}, sensitivity in position measurements \cite{Marquardts15}, or even bringing optomechanics close to the strong coupling regime with additional bath engineering \cite{Nori15}. In all these works, however, the nonlinear crystal is operated as a parametric amplifier, providing a nonlinear gain to some external field that is injected in the cavity at the down-converted frequency (\textit{stimulated} down-conversion). In contrast, the description of the interaction between the field generated via \cnb{\textit{spontaneous}} down-conversion and the mechanical mode is much more challenging, since (below threshold) the former is purely quantum mechanical \cite{cDOMPO}, so that the optomechanical coupling cannot be linearized and does not admit a simple Gaussian description.

In this work we provide a theory for the DOMPO which can be trusted all the way to threshold, and is obtained by combining traditional adiabatic elimination techniques with our recently developed self-consistent Mori projector (c-MoP) theory \cite{Degenfeld14,Degenfeld15}. To this end we first introduce the master equation which models the DOMPO, and perform an adiabatic elimination of the optical modes by neglecting the mechanical backaction. The mechanical state is found to stay approximately thermal for parameters compatible with current cWGM resonators, with an effective temperature dependent on the steady-state value and two-time correlation function of the down-converted photon number, which we derive in two ways. First by treating the pump mode as a classical field (semi-classical approach), allowing us to obtain simple analytical expressions \cnb{and provide a physical explanation for the regions of significant cooling, showing that the system provides a realistic implementation of the cooling-by-heating mechanism \cite{Mari12} below threshold}. Second, by using c-MoP theory on the optical dynamics to find reliable results at threshold and justify the absence of the mechanical backaction onto the optics. \cnb{Remarkably, this accurate approach allows us to prove that the semi-classical predictions break down \cnbb{when working very} close to threshold, where cooling is shown to disappear. \cnbb{These results might have strong implications not only for future analysis, but also for previous results which make use of semi-classical approaches while working very close to threshold \cite{Nori15}}.} In the final section we apply c-MoP theory on the full optomechanical problem, and identify the region of the parameter space where the mechanical backaction on the optics is negligible (which contains the parameters of our interest).

\section{The degenerate optomechanical parametric oscillator.}

\begin{figure}[t]
\includegraphics[width=\columnwidth]{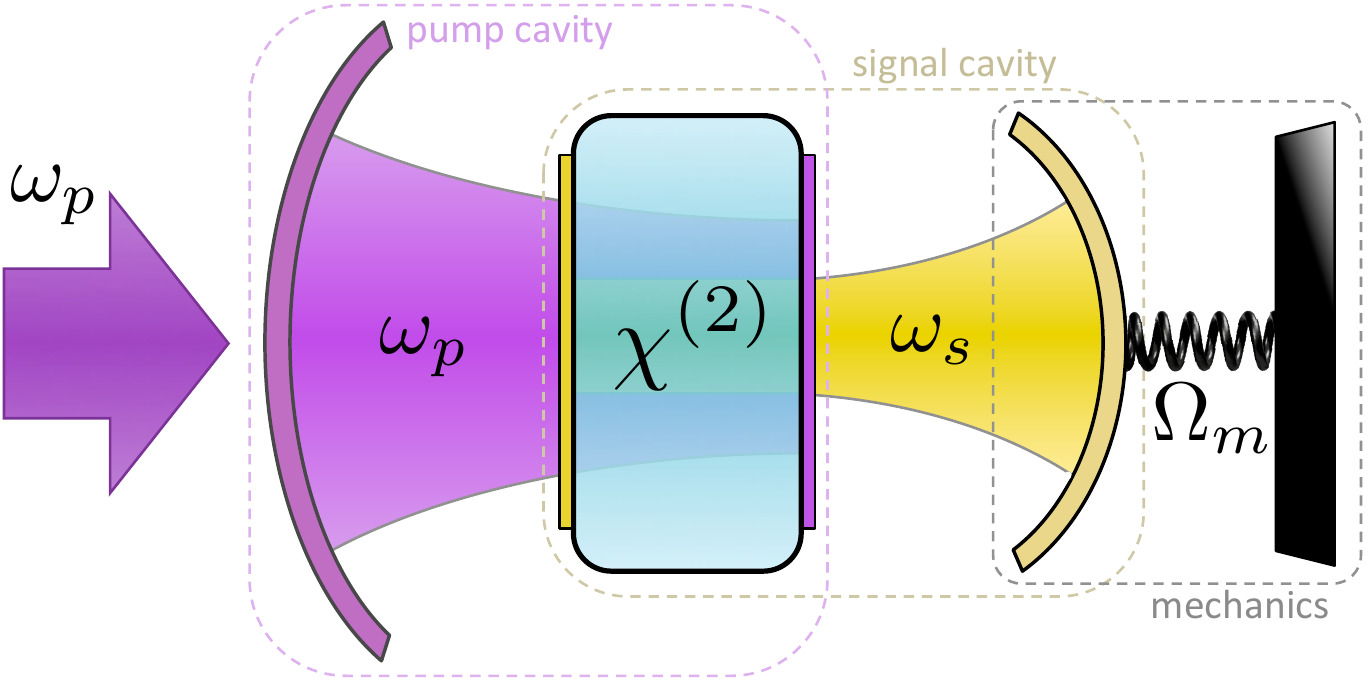}
\caption{\label{FigDOMPO} (Color online) Sketch of the degenerate optomechanical parametric oscillator. We take crystalline whispering gallery mode resonators \cite{Ilchenko03,Ilchenko04,Savchenkov07,Furst10,Furst10b,Hofer10,Furst11,Beckmann11,Werner12,Fortsch13,Marquardt13,Fortsch14,Fortsch14b} as a reference for our choice of parameters, but electromechanical implementations are also possible using superconducting circuits in the degenerate parametric oscillation configuration \cite{Leghtas15} coupled to drum-shaped oscillating capacitors \cite{Teufel11,Teufel11omit,Palomaki13} (see also \cite{Nation15}).}
\end{figure}

The system we consider can be schematically represented as in Fig. \ref{FigDOMPO}. A crystal with second-order optical nonlinearity is shared by two cavities with relevant resonances at frequencies $\omega_p$ (\textit{pump}) and $\omega_s\approx\omega_p/2$ (\textit{signal}). The pump cavity is driven by a resonant laser, so that photons in the signal cavity can be generated via \textit{spontaneous down-conversion} \cite{Boyd03,NavarretePhDthesis}. In addition, one of the mirrors of the signal cavity can oscillate, and is therefore \textit{optomechanically} coupled to the down-converted field via radiation pressure \cite{Aspelmeyer13}. Let us define annihilation operators $\{a_j\}_{j=p,s,m}$ for the pump ($p$), signal ($s$), and mechanical ($m$) modes. Including losses of the optical modes at rate $\gamma_0$ (assumed the same for pump and signal without loss of generality), as well as the irreversible energy exchange of the mechanical mode with its thermal environment at rate $\gamma_m$ (with which, in the absence of light, it is in thermal equilibrium with $\bar{n}_\mathrm{th}$ phonons), the master equation governing the evolution of the DOMPO's state $\rho$ can be written as
\begin{eqnarray} \label{masterDOMPO}
\dot\rho=-\mathrm{i}[H,\rho]&+&\gamma_0 \mathcal{D}_{a_s}[\rho]+\gamma_0 \mathcal{D}_{a_p}[\rho]
\\
&+&\gamma_m(\bar{n}_\mathrm{th}+1)\mathcal{D}_{a_m}[\rho]+ \gamma_m\bar{n}_\mathrm{th}\mathcal{D}_{a^\dagger_m}[\rho].\nonumber
\end{eqnarray}
We have defined Lindblad superoperators $\mathcal{D}_J[\cdot]=2J(\cdot) J^\dagger-J^\dagger J(\cdot)-(\cdot) J^\dagger J$, and the Hamiltonian
\begin{equation}\label{DOMPO Hamiltonian}
H=H_{\text{opt}}+\Omega_m a_m^\dagger a_m -\Omega_m\eta_\mathrm{OM}  a_s^\dagger a_s (a_m^\dagger+a_m),
\end{equation}
where we normalize the optomechanical coupling $\eta_\mathrm{OM}$ to the frequency of the mechanical oscillation $\Omega_m$. The optical Hamiltonian can be written in a picture rotating at the laser frequency as
\begin{equation}\label{optics Hamiltonian}
H_{\text{opt}}=\Delta_s a_s^\dagger a_s + \mathrm{i} \epsilon_p \left(a_p^\dagger-a_p \right) + \mathrm{i} \frac{\chi}{2} \left(a_p a_s^{\dagger \,2}-a_p^\dagger a_s^2 \right),
\end{equation}
where $\Delta_s=\omega_s-\omega_p/2$ is the detuning of the signal mode (which we will take positive in this work), $\chi/2$ is the down-conversion rate, and $\epsilon_p$ is proportional to the square root of the injected laser power.

In the classical limit, the steady-state phase diagram of the DOMPO features a variety of phases \cite{cDOMPO}. Here we focus on the regime where the state of the signal field is fully quantum, i.e. where the trivial solution $\langle a_s\rangle=0$ is the only stable one, henceforth referred to as the \emph{monostable phase}, which requires two conditions. First, defining the \emph{injection parameter} $\sigma=\epsilon_p\chi/\gamma_0^2$ and the normalized detuning $\Delta=\Delta_s/\gamma_0$, the trivial solution becomes unstable in favour of a nontrivial one $\langle a_s\rangle\neq0$ for $\sigma>\sqrt{1+\Delta^2}$ \cite{cDOMPO}. Hence, we write $\sigma=\sqrt{1+\Delta^2}\,x$ and focus on the $x\in [0,1]$ region. The second condition, $4 \Omega \Delta \eta_\mathrm{OM}^2/\eta_\mathrm{DC}^2<1$, guarantees that the nontrivial solution does not enter the $x\in [0,1]$ region \cite{cDOMPO}. In this expression we have introduced the dimensionless down-conversion coupling $\eta_\mathrm{DC}= \chi/ \gamma_0$ as well as the sideband-resolution parameter $\Omega=\Omega_m/\gamma_0$.

We emphasize that the vanishing signal field amplitude excludes the possibility of using a linearization approach similar to those applied in \cite{Agarwal09,Agarwal09bis,Marquardts15,Marquardt07,Genes08,Rips12,Navarrete14,Abdi15}. In the following, we provide a theory that works in all the $x\in[0,1]$ region, and use it to predict the action of the down-converted field on the mechanical state.

\section{Effective mechanical dynamics}

Despite the complexity of the problem, we remarkably find with the help of c-MoP theory that for typical system parameters the optical modes do not receive considerable backaction from the mechanics. This property, which we justify in Sec. \ref{BackAction}, allows us to simplify the problem significantly via an adiabatic elimination of the optical modes \cite{ZollerBook,BreuerPetruccione,CarmichaelBook1,WilsonRae07,Marquardt07} leading to an effective master equation for the reduced mechanical state $\rho_m(t)$. As we show in Appendix \ref{App1}, the mechanical steady state can then be approximated by a thermal state (displaced by $\langle a_m\rangle\approx\eta_\mathrm{OM}\bar{N}_s$) characterized by its phonon number
\begin{equation}\label{ss phonon number} 
\bar{n}_m=\frac{\bar n_\mathrm{th} +\Gamma_+}{1+(\Gamma_- - \Gamma_+)}\;\;{\overset{\Gamma\gg 1}{\approx}}\;\; \frac{\bar n_\mathrm{th}}{\Gamma}+ \bar{n}_{\text{FL}}\;,
\end{equation} 
where $\Gamma=\Gamma_--\Gamma_+$ is the cooling efficiency and $\bar{n}_\mathrm{FL} =\Gamma_+/\Gamma$ the fundamental limit for the phonon number. All the information about the optical modes is contained in the heating and cooling rates $\Gamma_\pm= C \operatorname{Re} \left\{\gamma_0\int_0^\infty d\tau \, \mathrm{exp}(\mp i \Omega_m \tau) \;s(\tau)\right\}$ through the optical correlation function
\begin{equation}\label{two time corr}
s(\tau)=\text{tr} \{a_s^\dagger a_s \,e^{\mathcal{L}_\mathrm{opt}\tau}a_s^\dagger a_s\,\bar{\rho}_\mathrm{opt}\}-\bar{N}_s^2,
\end{equation}
where $\bar{N}_s=\text{tr}\{a_s^\dagger a_s \bar{\rho}_\mathrm{opt}\}$ is the signal photon number and $C=\Omega Q\eta_\mathrm{OM}^2$ the bare cooperativity with $Q=\Omega_m/\gamma_m$ the mechanical quality factor. Here, $\bar{\rho}_\mathrm{opt}$ is the steady state of the optical Liouvillian
\begin{equation}\label{LoptNew}
\mathcal{L}_\mathrm{opt}[\cdot]=-\mathrm{i}[H_\mathrm{opt},\cdot]+\sum_{j={s,p}}\gamma_0 \mathcal{D}_{a_j}[\cdot],
\end{equation}
that is, $\mathcal{L}_{\text{opt}}[\bar{\rho}_\mathrm{opt}]=0$.

In the following we study the behaviour of the steady-state phonon number as we approach the DOMPO's threshold. From Eq. (\ref{ss phonon number}) it is clear that optimal cooling is then found by simultaneously maximizing $\Gamma$ and minimizing $\bar{n}_\mathrm{FL}$.

The nonlinear nature of the parametric down-conversion process in Eq. (\ref{LoptNew}) and a potential back action of the mechanical mode preclude an exact treatment of the optical correlation function in Eq. (\ref{two time corr}). To get simple analytic expressions that enable physical insight, we first apply standard linearization to the optical problem\cnb{, which we denote by \textit{semi-classical} approach and has been the method of choice in previous works \cite{Agarwal09,Agarwal09bis,Marquardts15,Nori15}}. \cnb{Next}, applying c-MoP theory \cite{Degenfeld15} we \cnb{show the failure} of the semi-classical approach close to the critical point and find more accurate expressions at criticality, which will also allow us to justify the adiabatic elimination of the optical modes.

\begin{figure}[t]
\includegraphics[width=\columnwidth]{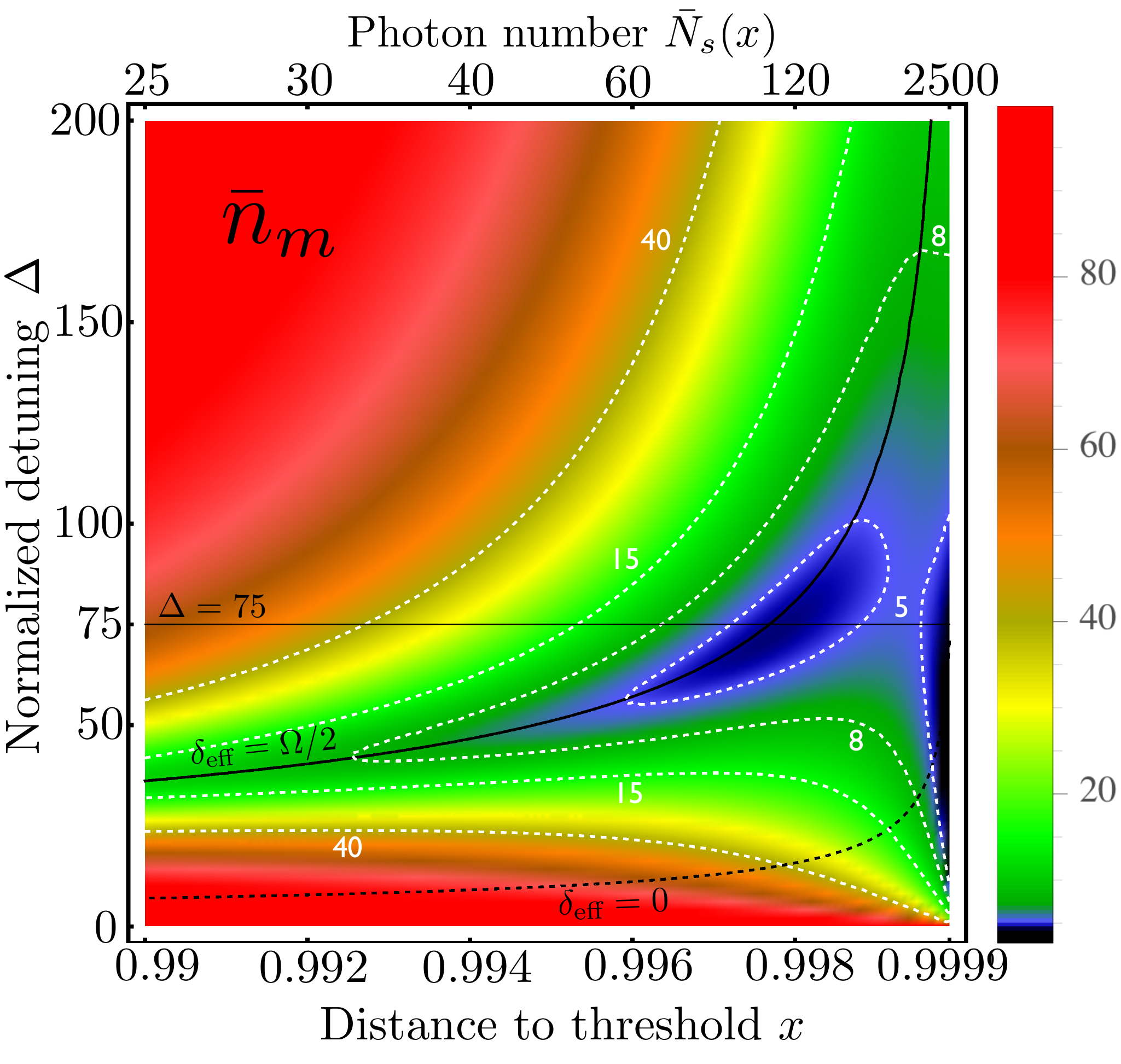}
\caption{\label{nf classical} (Color online) Steady-state phonon number as a function of $\Delta$ and $x$, as obtained from a semi-classical description of the DOPO and a thermal phonon number $\bar{n}_\mathrm{th}=100$ \cnb{(achievable either with cryogenics \cite{Riviere11} or via a standard sideband pre-cooling stage)}. We choose typical parameters for cWGM resonators \cnb{\cite{Hofer10}}: $\Omega=10$, $\eta_\mathrm{OM}=10^{-4}$, and $Q=10^6$. \cnbb{Note that typical optical decay rates are on the MHz range, although only normalized quantites are relevant for our results}.
The signal steady-state photon numbers $\bar{N}_s(x)$ corresponding to the ticked $x$-values are shown in the upper axis, showing that cooling is effective even with just $\sim 100$ photons. \cnb{On the other hand, c-MoP theory has allowed us to prove that the cooling region closer to threshold disappears once such a more accurate approach is used\cnbb{, see Fig. \ref{nf cMoP vs classical}.}}}
\end{figure}

\section{Semi-classical approach}

Below threshold, the linearization of the DOPO is accomplished by treating the pump mode as a classical stationary source, that is, by performing the replacement $a_p\rightarrow \epsilon_p/\gamma_0$ \cite{BlueBook,CarmichaelBook2,NavarretePhDthesis}. Within this approximation, the optical problem is governed by a Gaussian single-mode Liouvillian
\begin{equation}\label{Ls}
\gamma_0^{-1}\mathcal L_\mathrm{opt}[\cdot]=-\mathrm{i}[\Delta a_s^\dagger a_s+\sigma (a_s^{\dagger \,2}-a_s^2)/2,\cdot]+\mathcal{D}_{a_s}[\cdot],	
\end{equation}
from which any correlation function can be easily found, allowing us to obtain analytical expressions for the relevant quantities in Eq.~(\ref{ss phonon number}), as we show in Appendix \ref{Sec Semi-Class SupMat}. For the fundamental limit, we find
\begin{equation}
\bar{n}_{\text{FL}}=[4+(\Omega-2\Delta)^2]/8\Omega\Delta,
\end{equation}
while the cooling efficiency can be written as
\begin{equation}
\Gamma=Q\eta_\mathrm{OM}^2 \bar{N}_s(x)\Delta f(\Omega,\delta_\mathrm{eff}),
\end{equation}
where we have defined the function
\begin{equation}
f(\Omega,\delta_\mathrm{eff})=\frac{8 \Omega^2 [\Omega^2+4 (5+\delta_\mathrm{eff}^2)]}{(4+\Omega^2)[\Omega^4+16 (1+\delta_\mathrm{eff}^2)^2+8 \Omega^2 (1-\delta_\mathrm{eff}^2)]},\nonumber
\end{equation}
and a parameter $\delta_\mathrm{eff}=\sqrt{\Delta^2-\sigma^2}$ that will be shown later to play the important role of an effective optical detuning. The photon number $\bar{N}_s(x)=x^2/(2-2 x^2)$ is fully due to quantum fluctuations and increases hyperbolically until threshold $x=1$ where it diverges in this semi-classical approach.


In Fig.~\ref{nf classical} we show the steady-state phonon number as a function of the two control parameters, detuning $\Delta$ and distance to threshold $x$, fixing the rest of parameters to typical values of cWGM resonators \cite{Furst10b,Hofer10}. There are two regions where significant cooling effects appear. One is in the vicinity of the threshold point and can be traced back to the vast increase of the photon number $\bar{N}_s$ which makes $\Gamma\gg 1$ for virtually any value of the remaining parameters. However, as we will show below with the c-MoP approach, so close to threshold this semi-classical approach breaks down, \cnb{hence rendering this prediction incorrect}.

The other region, which turns out to be of major significance when aiming for optimal cooling, corresponds to $\delta_{\text{eff}}\approx \Omega/2$, see the black solid line in Fig.~\ref{nf classical}. The c-MoP approach will confirm this prediction in the next section. Moreover, it can be understood in physical terms by moving to a new picture defined by the squeezing operator $S(r)=\exp[-\mathrm{i}r(a_s^{\dagger 2}+a_s^2)/2]$ with $\tanh 2r=\sigma/\Delta$ (note that such transformation requires $\Delta>\sigma$, which corresponds in Fig.~\ref{nf classical} to the region above the black dashed line). \cnb{This transformation diagonalizes the Hamiltonian in the optical Liouvillian (\ref{Ls}), so that, defining the parameters $\bar{N}_\mathrm{eff}=(\Delta/\delta_{\text{eff}}-1)/2$ and $M=\sigma/2\delta_{\text{eff}}$, the transformed state $\tilde{\rho}=S^\dagger(r)\rho S(r)$ evolves according to
\begin{eqnarray}\label{NewPictureME}
&&\gamma_0^{-1}\partial_t\tilde{\rho}=\left[-\mathrm{i}\delta_\mathrm{eff} a_s^\dagger a_s,\tilde{\rho}\right]+(\bar{N}_\mathrm{eff}+1) \mathcal{D}_{a_s}[\tilde{\rho}]+\bar{N}_\mathrm{eff}\mathcal{D}_{a_s^\dagger}[\tilde{\rho}] \nonumber
\\
&&-[\mathrm{i}\Omega a_m^\dagger a_m,\tilde{\rho}]+(\Omega/Q)\{(\bar{n}_\mathrm{th}+1)\mathcal{D}_{a_m}[\tilde{\rho}]+\bar{n}_\mathrm{th}\mathcal{D}_{a_m^\dagger}[\tilde{\rho}]\} \nonumber
\\
&&+ [\Omega\eta_\mathrm{OM}M(a_s^2 a_m^\dagger-a_s^{\dagger 2} a_m),\tilde{\rho}],
\end{eqnarray}
within the rotating-wave approximation, valid under the conditions $4\delta_\text{eff}^2\gg \sigma$ and $\eta_\text{OM}\Delta\ll \delta_\text{eff}$, see Appendix \ref{cooling by heating App}.}

Therefore, in this picture the signal field is turned into a bosonic mode with oscillation frequency $\delta_\mathrm{eff}$ and thermal occupation $\bar{N}_\mathrm{eff}$. The optomechanical coupling is dressed by the squeezing parameter $M$, similarly to the dressing by the intracavity field amplitude in standard sideband cooling \cite{WilsonRae07,Marquardt07}. However, at difference with that case the interaction exchanges phonons with pairs of photons (rather than single photons), thus explaining why $\delta_\mathrm{eff}=\Omega/2$ is the resonance condition for cooling. Within this condition, assuming $2\bar{N}_\mathrm{eff}\gg 1$ and $\Omega^2\gg 4$, we furthermore find $\Gamma\approx 2C M^2 \bar{N}_\mathrm{eff}$. The cooling efficiency $\Gamma$ thus receives an additional contribution $2\bar{N}_\mathrm{eff}$ from the effective thermal photon number, which is a direct consequence of the nonlinear nature of the effective optomechanical coupling in (\ref{NewPictureME}) that cannot be found in standard sideband cooling, as we discuss in Appendix \ref{cooling by heating App}. This represents a natural example of the so-called \textit{cooling-by-heating} effect \cite{Mari12}, where heating up the optical field can contribute to making optomechanical cooling more efficient. However, as it is well known from standard sideband cooling \cite{Dobrindt08}, thermal photons also contribute to the fundamental limit, which indeed can be approximated by $\bar{n}_\mathrm{FL}\approx \bar{N}_\mathrm{eff}/2$ in \cnb{our} scenario. When the term $\bar{n}_\mathrm{th}/\Gamma$ dominates over $\bar{n}_\mathrm{FL}$ in Eq.~(\ref{ss phonon number}) the thermal optical background $\bar{N}_\mathrm{eff}$ can then be interpreted as ``good noise'', while as soon as the fundamental limit is reached it becomes ``bad noise'' and heats up the mechanical motion (see the thick black curve in Fig. \ref{nf classical}). We discuss this phenomenon in more detail in Appendix \ref{cooling by heating App}.

\begin{figure}[t]
\includegraphics[width=\columnwidth]{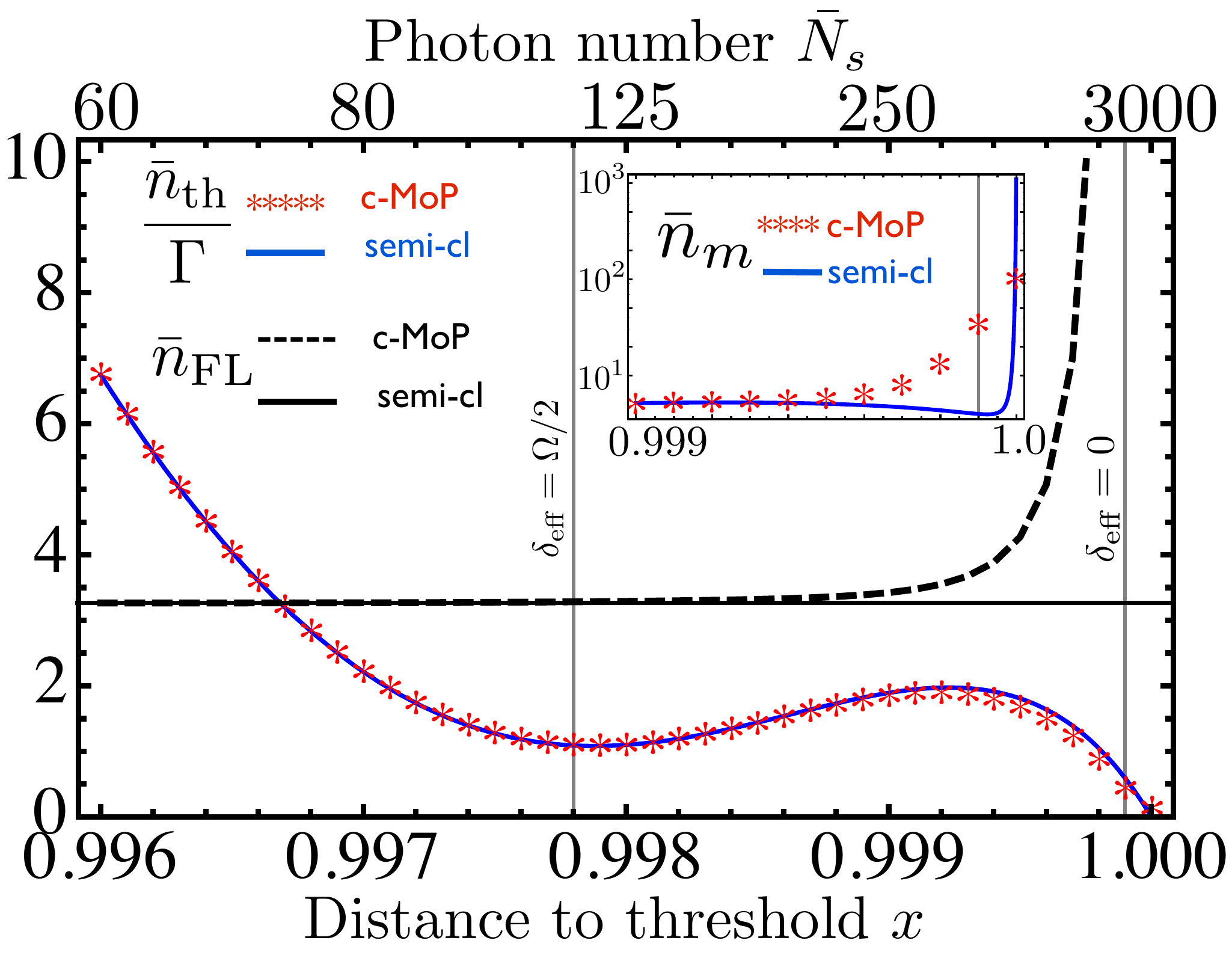}
\caption{\label{nf cMoP vs classical} (Color online) Terms $\bar{n}_\mathrm{th}/\Gamma$ and $\bar{n}_\mathrm{FL}$ contributing to the steady-state phonon number (\ref{ss phonon number}) as a function of the \cnb{distance-to-threshold} parameter $x$. We fix the detuning to $\Delta=75$, corresponding to the straight thin solid line in Fig. \ref{nf classical}, taking $\eta_\text{DC}=0.01$ \cite{Furst10b} and the rest of parameters as in that figure. The inset displays a close up of the steady-state phonon number $\bar{n}_m$ for $x\in[0.999,1]$, but without applying the rotating-wave and Markov approximations, see \ref{mech moments}. Note that c-MoP gives finite results equivalent to those found within the rotating-wave approximation, while the semi-classical predictions diverge at threshold, which can be taken as further evidence that the theory breaks down there.
}
\end{figure}

\cnb{It is interesting to examine the limits that this ``bad noise'' imposes on cooling. For the parameters of Fig. \ref{nf cMoP vs classical}, compatible with current cWGM resonators, it is found that the cooling efficiency and the fundamental limit can be simultaneously optimized to about $\Gamma\approx 85$ and $\bar{n}_\text{FL}\approx 3$ as a function of the experimentally-tunable parameters $\Delta$ and $\sigma$, precluding ground-state cooling. However, the foreseeable increase of the mechanical quality factor and the optomechanical coupling by one order of magnitude in next-generation cWGM resonators \cite{Hofer10} will improve these numbers to $\Gamma\approx 400$ and $\bar{n}_\text{FL}\approx 0.15$.}

\section{c-MoP approach}

The semi-classical approach has allowed us to get analytical and physical insight into the problem. It is however well known that this approximation fails close to the critical point\cnb{, although there is no systematic way of checking where exactly within the semi-classical formalism itself. Hence,} to determine where it exactly breaks down and to find more accurate results for those parameters, we make use of the recently developed c-MoP technique \cite{Degenfeld14,Degenfeld15}, which allows finding reduced equations for the constituent parts of a composite system, even in situations where there is significant backaction among its parts and no time-scale separation between their dynamics.

For parameters compatible with cWGM resonators, the theory is already regularized by using c-MoP only in the optical problem (DOPO), which provides a more accurate description for the optical correlation function (\ref{two time corr}) and photon number that enter the effective mechanical dynamics. The application of c-MoP to the DOPO has been detailed in \cite{Degenfeld15}, but we review its most relevant steps for completeness in Appendix \ref{c-MoP theory}. Specifically, we use a combination of c-MoP and a Gaussian-state approximation, which provides an efficient and accurate tool capable of regularizing the divergencies and unphysical predictions of the semi-classical approach. In particular, we show in Appendix \ref{c-MoP theory} that at threshold the decay rate of the optical correlator scales as $\gamma_\mathrm{opt}\propto\gamma_0\eta_\mathrm{DC}(1+\Delta)$, while the photon number as $\bar{N}_s\propto(1+\Delta)/\eta_\mathrm{DC}$, in contrast to semi-classical results in which the former goes to zero while the latter diverges.

We show a very representative case for the phonon number $\bar{n}_m$ as a function of the distance to threshold $x$ in Fig.~\ref{nf cMoP vs classical}. The method finds perfect agreement with the semi-classical predictions \cnb{sufficiently below threshold, in particular} verifying the cooling-by-heating effect presented above. Most importantly, we find that the absolute minimum phonon number is indeed reached when the resonance condition $\delta_{\text{eff}}=\Omega/2$ is met. On the other hand, close to threshold we find a significant correction to the semi-classical predictions for the fundamental limit $\bar{n}_\mathrm{FL}$. In particular, while this is independent of the distance to threshold $x$ in the semi-classical picture, c-MoP shows that it actually increases very rapidly as the critical point is approached, and hence no cooling is found no matter how much the efficiency $\Gamma$ is increased. This is consistent with the fact that when $\Delta<\sigma$ (as happens at threshold), $\delta_\mathrm{eff}$ becomes imaginary and there is no resonance for the optomechanical interaction.

\section{Absence of mechanical backaction onto the optics}\label{BackAction}

The adiabatic elimination of the optical fields which we have used throughout the work relies on the time-scale separation between the optical and mechanical degrees of freedom. In particular, such an approach neglects mechanical backaction onto the optics, which is a good approximation as long as the rate of any mechanical perturbation is much smaller than the intrinsic relaxation rate of the optics $\gamma_\mathrm{opt}$. Far from the critical point the optical relaxation rate is $\gamma_0$, which usually dominates over any other rate in the system. However, as the critical point is approached the DOPO dynamics exhibits a critical slowing down, and its relaxation rate becomes smaller and smaller. Hence, in our work that considers parameters close to threshold, it is very important to check that the desired time-scale separation is present.

An intuitive argument supporting such a time-scale separation follows from relating the mechanical backaction rate with the optical frequency-shift induced by the optomechanical interaction, $\gamma_\mathrm{back}=\eta_\mathrm{OM}\Omega_m \langle x_m\rangle=2\eta_\mathrm{OM}^2\Omega_m\bar{N}_s$, where we have used (\ref{displacement}). Hence, using the scaling of $\bar{N}_s$ and $\gamma_\mathrm{opt}$ obtained in the previous section at threshold, the condition $\gamma_\mathrm{back}\ll \gamma_\mathrm{opt}$ becomes $2\Omega(\eta_\mathrm{OM}/\eta_\mathrm{DC})^2\ll 1$, which is very well satisfied for the parameters that we work with. Moreover, note that this condition is automatically satisfied when working within the monostability condition $4\Delta\Omega(\eta_\mathrm{OM}/\eta_\mathrm{DC})^2<1$ as long as $\Delta\gg 1/2$.

We can set more rigorous bounds to the region where mechanical backaction is negligible by using c-MoP theory \cite{Degenfeld14,Degenfeld15}, since, in contrast to adiabatic elimination methods, it does not rely on the concept of time-scale separation or absence of backaction effects. Hence, we apply this theory to the DOMPO system by using the time-dependent self-consistent Mori projectors $\mathcal{P}_t^\mathrm{opt}=\rho_\mathrm{opt}(t)\otimes\mathrm{tr}_\mathrm{opt}\{\cdot\}$ and $\mathcal{P}_t^m=\mathrm{tr}_m\{\cdot\}\otimes\rho_m(t)$, that is, using a bipartition ``optics $\otimes$ mechanics'' for the system. This approach will allow us to identify the terms contributing to the mechanical backaction and find upper bounds to their scaling. Following the procedure introduced in previous works \cite{Degenfeld14,Degenfeld15}, the c-MoP equations for the reduced optical and mechanical states in the asymptotic $t\rightarrow\infty$ or steady-state limit are easily found to read
\begin{subequations}\label{cMoPforDOMPO}
\begin{align}
&\frac{d\bar{\rho}_m}{dt}=0=\mathcal{L}_m\bar{\rho}_m+\mathrm{i}\Omega_m\eta_\mathrm{OM}\langle a_s^\dagger a_s\rangle[x_m,\bar{\rho}_m]\label{cMoP for m}
\\
&-\Omega_m^2\eta_\mathrm{OM}^2\left[x_m,\int_0^\infty d\tau \,e^{\mathcal{L}_m\tau}(\delta x_m\bar{\rho}_m s(\tau)-\mathrm{H.c.})\right], \nonumber
\\
&\frac{d\bar{\rho}_\mathrm{opt}}{dt}=0=\mathcal{L}_\mathrm{opt}\bar{\rho}_\mathrm{opt}+\mathrm{i}\Omega_m\eta_\mathrm{OM} \langle x_m\rangle[a_s^\dagger a_s,\bar{\rho}_\mathrm{opt}] \label{cMoP for optics}
\\
&-\Omega_m^2\eta_\mathrm{OM}^2\left[a_s^\dagger a_s,\int_0^\infty \hspace{-2mm}d\tau \,e^{\mathcal{L}_\mathrm{opt}\tau}(\delta n_s\bar{\rho}_\mathrm{opt}\, s_m(\tau)-\mathrm{H.c.})\right],\nonumber
\end{align}	
\end{subequations}
where, for any operator $A$, we have introduced the usual definitions $\langle A\rangle=\mathrm{tr}\{A(\bar{\rho}_s\otimes\bar{\rho}_p)\}$ and $\delta A=A-\langle A\rangle$, while $\delta n_s=a_s^\dagger a_s-\langle a_s^\dagger a_s\rangle$. $s(\tau)$ is the usual optical correlation function (\ref{two time corr}), and we have defined the mechanical correlation function
\begin{align}\label{m corr func}
s_m(\tau)&=\mathrm{tr}_m\{x_m e^{\mathcal{L}_m\tau} \delta x_m\bar{\rho}_m\}=e^{(-\mathrm{i}\Omega_m -\gamma_m)\tau}
\\
&+\left[e^{(-\mathrm{i}\Omega_m-\gamma_m)\tau}(\langle\delta a_m^\dagger \delta a_m\rangle+\langle \delta a_m^2\rangle)+\text{H.c.}\right],\nonumber
\end{align}
where the final expression is easily found by following a similar procedure as the one shown in Appendix \ref{Sec Semi-Class SupMat} for the semi-classical optical correlation function, since $\mathcal{L}_m$ is quadratic.

The last two terms on the right-hand side of Eq. (\ref{cMoP for optics}) account for the mechanical backaction on the optics, and we proceed now to bound their effect. The second to last term describes precisely the optical detuning $\Omega_m\eta_\mathrm{OM}\langle x_m\rangle$ induced by the optomechanical interaction that we already discussed at the beginning of the section, concluding that it is negligible within the parameter regime we work with. Then we focus on the last term in Eq. (\ref{cMoP for optics}), the Born term. For this purpose we derive the steady-state equation for the moment $\langle a_s^{\dagger 2}\rangle$ (note that the equation of motion of $\langle a^\dagger_s a_s\rangle$ receives no explicit mechanical backaction even within c-MoP theory) which reads
\begin{align}\label{as2 eq}
&0= \mathrm{tr}_\mathrm{opt}\{a_s^{\dagger 2} \mathcal{L}_\mathrm{opt}\bar{\rho}_\mathrm{opt}\}-2\mathrm{i}\Omega_m\eta_\mathrm{OM} \langle x_m\rangle \langle a_s^{\dagger 2}\rangle
\\
&\hspace{-1mm}+\hspace{-0.5mm}2\Omega_m^2\eta_\mathrm{OM}^2 \int_0^\infty \hspace{-2mm}d\tau \,\mathrm{Re}\{s_m(\tau)\} \,\mathrm{tr}_\mathrm{opt}\left\{a_s^{\dagger 2}e^{\mathcal{L}_\mathrm{opt}\tau}[a_s^\dagger a_s,\bar{\rho}_\mathrm{opt}]\right\} \nonumber
\\
&\hspace{-1mm}+\hspace{-0.5mm}2\mathrm{i}\Omega_m^2\eta_\mathrm{OM}^2 \int_0^\infty \hspace{-3mm}d\tau \,\mathrm{Im}\{s_m(\tau)\} \,\mathrm{tr}_\mathrm{opt}\left\{a_s^{\dagger 2}e^{\mathcal{L}_\mathrm{opt}\tau}\{\delta n_s,\bar{\rho}_\mathrm{opt}\}\right\}\hspace{-1mm}, \nonumber
\end{align}
where $\{\cdot,\cdot \}$ denotes the anticommutator. Note first that the correlation functions $\mathrm{tr}_\mathrm{opt}\left\{a_s^{\dagger 2}e^{\mathcal{L}_\mathrm{opt}\tau}[a_s^\dagger a_s,\bar{\rho}_\mathrm{opt}]\right\}$
and $\mathrm{tr}_\mathrm{opt}\left\{a_s^{\dagger 2}e^{\mathcal{L}_\mathrm{opt}\tau}\{\delta n_s,\bar{\rho}_\mathrm{opt}\}\right\}$, which are of similar structure as the optical correlation function $s(\tau)$, decay to zero at a rate $\gamma_{\text{opt}}$. Next, we derive upper bounds for the last two terms in Eq. (\ref{as2 eq}). For the second to last term we find
\begin{align}\label{estimate1}
&\left| 2\Omega_m^2\eta_\mathrm{OM}^2 \int_0^\infty d\tau \,\mathrm{Re}\{s_m(\tau)\} \,\mathrm{tr}_\mathrm{opt}\left\{a_s^{\dagger 2}e^{\mathcal{L}_\mathrm{opt}\tau}[a_s^\dagger a_s,\bar{\rho}_\mathrm{opt}]\right\}\right| \nonumber
\\
&\leq \left| 2\Omega_m^2\eta_\mathrm{OM}^2 \frac{\mathrm{Re}\{s_m(0)\} \,\mathrm{tr}_\mathrm{opt}\left\{a_s^{\dagger 2}e^{\mathcal{L}_\mathrm{opt}0}[a_s^\dagger a_s,\bar{\rho}_\mathrm{opt}]\right\}}{\gamma_m+\gamma_\mathrm{opt}}\right| \nonumber
\\
&\leq \underbrace{\frac{4\Omega_m^2\eta_\mathrm{OM}^2\bar{n}_m}{\gamma_m+\gamma_\mathrm{opt}}}_{\gamma_\mathrm{back}^\prime} |\langle a_s^{\dagger 2}\rangle|,
\end{align}
where in the last step we have used
$s_m(0)=1+\langle\delta a_m^\dagger\delta a_m\rangle+\langle\delta a_m^2\rangle\approx\langle\delta a_m^\dagger\delta a_m\rangle\equiv\bar{n}_m$ (note that we expect the mechanical state to stay approximately thermal, and hence $\langle\delta a_m^2\rangle\approx 0$) and $\mathrm{tr}_\mathrm{opt}\left\{a_s^{\dagger 2}[a_s^\dagger a_s,\bar{\rho}_\mathrm{opt}]\right\}=\langle[a_s^{\dagger 2},a_s^\dagger a_s]\rangle=2\langle a_s^{\dagger 2}\rangle$. Similarly, for the last term in Eq. (\ref{as2 eq}) we find
\begin{align}\label{estimate}
\left|2\mathrm{i}\Omega_m^2\eta_\mathrm{OM}^2\right.&\left.\int_0^\infty \hspace{-2mm}d\tau \,\mathrm{Im}\{s_m(\tau)\} \,\mathrm{tr}_\mathrm{opt}\left\{a_s^{\dagger 2}e^{\mathcal{L}_\mathrm{opt}\tau}\{\delta n_s,\bar{\rho}_\mathrm{opt}\}\right\}\right|\nonumber
\\
&\leq\left|2\Omega_m^2\eta_\mathrm{OM}^2 \frac{\mathrm{tr}_\mathrm{opt}\left\{a_s^{\dagger 2}e^{\mathcal{L}_\mathrm{opt}0}\{\delta n_s,\bar{\rho}_\mathrm{opt}\}\right\}}{\gamma_m+\gamma_\mathrm{opt}}\right|\nonumber
\\
&\leq \frac{2\Omega_m^2\eta_\mathrm{OM}^2}{\gamma_m+\gamma_\mathrm{opt}} \left|\mathrm{tr}_\mathrm{opt}\left\{a_s^{\dagger 2}\{\delta n_s,\bar{\rho}_\mathrm{opt}\}\right\}\right|\nonumber
\\
&\approx \underbrace{\frac{4\Omega_m^2\eta_\mathrm{OM}^2(2\bar{N}_s +1)}{\gamma_m+\gamma_\mathrm{opt}}}_{\gamma_\mathrm{back}^{\prime\prime}} |\langle a_s^{\dagger 2}\rangle|.
\end{align}
where for the last expression we have used $\mathrm{tr}_\mathrm{opt}\left\{a_s^{\dagger 2}\{\delta n_s,\bar{\rho}_\mathrm{opt}\}\right\}=2 (\langle a_s^{\dagger 3} a_s\rangle -\langle a_s^{\dagger 2}\rangle \langle a_s^\dagger a_s\rangle+\langle a_s^{\dagger 2}\rangle) \approx 2(2\bar{N}_s +1)\langle a_s^{\dagger 2}\rangle$, within the Gaussian state approximation (\ref{GSAmoments}), that is, $\langle a_s^{\dagger 3} a_s\rangle\approx 3 \langle a_s^{\dagger 2}\rangle \langle a_s^\dagger a_s\rangle$, noting that $\langle a_s\rangle=0$ below threshold.

A sufficient condition for mechanical backaction to be negligible is then $\gamma_\mathrm{back}^\prime,\gamma_\mathrm{back}^{\prime\prime}\ll\gamma_\mathrm{opt}$. We pass to check whether this is the case in our work. Note first that $\gamma_m\ll\gamma_\mathrm{opt}$ even at threshold, since $\gamma_\mathrm{opt}/\gamma_m\sim\gamma_0\eta_\mathrm{DC}(1+\Delta)/\gamma_m\gg 1$ for the parameters we are interested in. Using the scalings $\gamma_\mathrm{opt}\propto\gamma_0\eta_\mathrm{DC}(1+\Delta)$ and $\bar{N}_s\propto(1+\Delta)/\eta_\mathrm{DC}$ at threshold (where these bounds are the tightest), we can then write the conditions under which backaction is negligible as
\begin{subequations}
\begin{eqnarray}
\frac{\gamma_\mathrm{back}^\prime}{\gamma_\mathrm{opt}} &\sim & \frac{\Omega^2\bar{n}_m\eta_\mathrm{OM}^2}{\eta_\mathrm{DC}^2(1+\Delta)^2}\ll 1,
\\
\frac{\gamma_\mathrm{back}^{\prime\prime}}{\gamma_\mathrm{opt}} &\sim & \frac{\Omega^2\eta_\mathrm{OM}^2}{\eta_\mathrm{DC}^3(1+\Delta)}\ll 1.
\end{eqnarray}
\end{subequations}
For the parameter set of Fig. 2 these lead to the conditions $\bar{n}_m\ll 100(1+\Delta)^2$ and $1+\Delta\gg 1$, respectively. For the large values of $\Delta$ that we use during most of the work, these conditions are very well satisfied. For small $\Delta$ they seem to be too tight, but we need to stress here that we have been extremely conservative when estimating the Born terms (\ref{estimate1}) and (\ref{estimate}), meaning that in practice backaction should be negligible even in a much broader region of the parameter space.

Overall, c-MoP theory has allowed us to quantify the mechanical backaction on the optics in a rigorous manner. We have obtained very conservative bounds that the system parameters must satisfy in order for such backaction to be negligible, showing that this is indeed the case for the parameters used in our work, which are compatible with an implementation based on cWGM resonators. It is however foreseeable that such devices, as well as their electromechanical counterparts, will be able to study regions where backaction is significant, in which case the c-MoP approach presented in this section will be very useful.

\section{Conclusions}

By exploiting adiabatic elimination techniques, semi-classical methods, and c-MoP theory, we have provided a theoretical analysis of the DOMPO which works even at the critical point. We have focused on the region where the optical field is fully quantum, showing that such a quantum-correlated field with no coherent component can induce significant mechanical cooling through a cooling-by-heating mechanism. c-MoP techniques have allowed us to check the validity of the optical adiabatic elimination as well as the semi-classical approximation, whose predictions have indeed been shown to break down at threshold, showing the potential of c-MoP to treat dissipative quantum-optical problems in the vicinity of critical points.
%


\acknowledgments
We thank Yue Chang, Tao Shi, Alessandro Farace, Eugenio Rold\'an, Germ\'an J. de Valc\'arcel, Florian Marquardt, Christoph Marquardt, and J. Ignacio Cirac for useful discussions and comments. This work has been supported by the German Research Foundation (DFG) via the CRC 631 and the grant HA 5593/3-1. M.A. and C.N.-B. acknowledge funding from the Alexander von Humbolt Foundation through their Fellowship for Postdoctoral Researchers. M.J.H. acknowledges support by UK Engineering and Physical Sciences Research Council (EPSRC) under EP/N009428/1.

\appendix

\section{Elimination of the optical modes}\label{App1}

Here we present a derivation of the effective mechanical master equation leading to the phonon number of Eq.~(4) in the main text. Our starting point is the master equation governing the evolution of the state $\rho(t)$ of the DOMPO, Eq. (1), which for convenience we rewrite here as
\begin{equation}\label{App full master equation}
\partial_t \rho = \mathcal{L}_\mathrm{opt}[\rho]+\mathcal{L}_m[\rho]+\mathcal{L}_\mathrm{OM}[\rho]
\end{equation}
with
\begin{subequations}
\begin{align}
\mathcal{L}_\mathrm{opt}[\rho] &= \left[-\mathrm{i}\Delta_s a_s^\dagger a_s + \left(\epsilon_p a_p^\dagger + \frac{\chi}{2}a_p a_s^{\dagger \,2}-\text{H.c.}\right),\rho\right] \nonumber
\\
&+\gamma_0\mathcal{D}_{a_p}[\rho]+\gamma_0\mathcal{D}_{a_s}[\rho],
\\
\mathcal{L}_m[\rho] &=[-\mathrm{i}\Omega_m a_m^\dagger a_m,\rho] \nonumber
\\
&+\gamma_m (\bar{n}_\mathrm{th}+1)\mathcal{D}_{a_m}[\rho]+\gamma_m \bar{n}_\mathrm{th}\mathcal{D}_{a_m^\dagger}[\rho],
\\
\mathcal{L}_\mathrm{OM}[\rho] &= [\mathrm{i}\Omega_m\eta_\mathrm{OM} a_s^\dagger a_s (a_m+a_m^\dagger),\rho].
\end{align}
\end{subequations}
All the quantities are defined in the main text, and we remind the notation $\mathcal{D}_J[\rho]=2J\rho J^\dagger-J^\dagger J\rho-\rho J^\dagger J$ for superoperators in Lindblad form.

In order to eliminate the optical modes and find an effective master equation for the mechanical state $\rho_m(t)$ we proceed as follows. We first define  the projector superoperator $\mathcal{P}[\cdot]=\bar{\rho}_\mathrm{opt}\otimes \mathrm{tr}_\mathrm{opt}\{\cdot\}$ whose action on the full state $\rho(t)$ of the DOMPO is $\mathcal{P}[\rho(t)]=\bar{\rho}_\mathrm{opt}\otimes\rho_m(t)$. Here, $\bar{\rho}_\mathrm{opt}$ is the steady state of the optical Liouvillian, that is, $\mathcal{L}_\mathrm{opt}[\bar{\rho}_\mathrm{opt}]=0$. Applying this superoperator and its complement $1-\mathcal{P}$ onto the master equation, and formally integrating the latter, we obtain an exact equation of motion for $\rho_m(t)$, the so-called \emph{Nakajima-Zwanzig} equation \cite{BreuerPetruccione,ZollerBook}. Such equation is not solvable, and therefore we apply a \emph{Born approximation} which takes into account terms up to second order in the optomechanical interaction. The resulting equation reads
\begin{align}\label{NZE2}
&\dot \rho_m(t)=\mathcal L_m \rho_m(t)+\mathrm{i} \Omega_m\eta_\mathrm{OM} \bar{N}_s \left[x_m,\rho_m(t)\right]
\\
&-\Omega_m^2 \eta_\mathrm{OM}^2 \left[x_m,\int_0^t d\tau \,e^{\mathcal{L}_m\tau} [x_m \rho_m(t-\tau) s(\tau)-\mathrm{H.c.}]\right], \nonumber
\end{align}
where we have defined the mechanical position quadrature $x_m=a_m+a_m^\dagger$, the photon number in the signal mode $\bar{N}_s = \mathrm{tr}\{a_s^\dagger a_s \bar{\rho}_\mathrm{opt}\}$, and the optical correlation function
\begin{equation}\label{s corr func}
s(\tau)=\mathrm{tr}\{a_s^\dagger a_s e^{\mathcal L_\mathrm{opt}\tau}[a_s^\dagger a_s\bar{\rho}_\mathrm{opt}]\}-\bar{N}_s^2.
\end{equation}

It is well known that the steady state $\bar{\rho}_\mathrm{opt}$ of the DOPO is unique (which intuitively comes from the fact that both the pump and signal modes have local dynamics leading to unique steady states, and the parametric interaction preserves that uniqueness), and hence $e^{\mathcal{L}_\mathrm{opt}\tau}$ is a \emph{relaxing map} \cite{Rivas12,Schirmer}, mapping all optical operators $O$ into the steady state, that is, $\lim_{\tau\to\infty}e^{\mathcal{L}_\mathrm{opt}\tau}[O]=\mathrm{tr}_\mathrm{opt}\{O\}\bar{\rho}_\mathrm{opt}$. Thus, the optical correlation function $s(\tau)$ will always decay to zero within some finite \emph{memory time} which we denote by $\tau_\mathrm{opt}$. Hence, in the asymptotic limit we can write $\lim_{t\to\infty}\rho_m(t-\tau)=\lim_{t\to\infty}\rho_m(t)\equiv\bar{\rho}_m$ in the integral Kernel of Eq.~(\ref{NZE2}), obtaining an equation for $\bar{\rho}_m$ which is quadratic in the operators $a_m$ and therefore allows for a Gaussian-state solution \cite{Weedbrook12,NavarreteQInotes}. In other words, the equations for the first and second steady-state mechanical moments form a closed linear algebraic set
\begin{subequations}\label{mech moments}
\begin{eqnarray}
0&&=(-\mathrm{i} \Omega_m-\gamma_m) \langle a_m\rangle+\mathrm{i}\Omega_m\eta_\mathrm{OM}\bar{N}_s
\\
&&- \Omega_m^2\eta_\text{OM}^2\mathrm{Re}\{d_0\}\langle x_m\rangle,\nonumber
\\
0&&=\gamma_m (\bar{n}_\mathrm{th}-\langle\delta a_m^\dagger \delta a_m\rangle)
\\
&&\hspace{-1.5mm}-\hspace{-0.5mm} \Omega_m^2\eta_\mathrm{OM} \mathrm{Re}\{(d_+\hspace{-0.5mm}-\hspace{-0.5mm}d_-)\langle \delta a_m^\dagger \delta a_m\rangle\hspace{-0.5mm}+\hspace{-0.5mm}(d_-^*\hspace{-0.5mm}-\hspace{-0.5mm}d_+)\langle\delta a_m^2\rangle \hspace{-0.5mm}-\hspace{-0.5mm} d_-\},\nonumber
\\
0&&=(-\mathrm{i} \Omega_m-\gamma_m)\langle\delta a_m^2\rangle
\\
&&\hspace{-1mm}-\Omega_m^2\eta_\mathrm{OM}^2[(d_--d_+^*)\langle \delta a_m^\dagger \delta a_m\rangle+(d_+-d_-^*)\langle\delta a_m^2\rangle+d_-],\nonumber
\end{eqnarray}
\end{subequations}
where we used the abbreviations $\langle A\rangle=\mathrm{tr}\{A\bar{\rho}_m\}$, $\delta A=A-\langle A\rangle$, and
\begin{subequations}\label{d1 and d2}
\begin{align}
d_0 &= \int_0^\infty d\tau s(\tau),
\\
d_\pm &= \int_0^\infty d\tau \, e^{(\pm\mathrm{i}\Omega_m-\gamma_m)\tau} \,s(\tau).
\end{align}
\end{subequations}

These equations can be solved for the steady-state moments as functions of the optical photon number $\bar{N}_s$ and correlation function $s(\tau)$ without the need of further approximations. However, in order to obtain more physical insight into the mechanical steady state $\bar{\rho}_m$ we apply both Markov approximation and a \emph{rotating-wave approximation} to Eq. (\ref{NZE2}). The Markov approximation is based on the assumption that within the optical memory time $\tau_\mathrm{opt}$ all the mechanical dynamics can be neglected except for the evolution provided by the free Hamiltonian $\Omega_m a_m^\dagger a_m$. As a result we can write $e^{\mathcal{L}_m\tau}[x_m\rho_m(t-\tau)]\approx x_m(\tau) \rho_m(t)$ with $x_m(\tau)=e^{i\Omega_m\tau} a_m+e^{-i\Omega_m\tau} a_m^\dagger$. On the other hand, the rotating-wave approximation consists in neglecting all the terms proportional to $a_m^2$ and $a_m^{\dagger 2}$ in the effective mechanical master equation, under the assumption that their rotation at frequency $2\Omega_m$ is much larger than the rates they are weighted by. After applying these approximations in Eq.~(\ref{NZE2}) we are left with an effective mechanical master equation given by
\begin{align}\label{ME Born-Markov plus RWA}
\dot\rho_m&=\mathcal L_m \rho_m +\mathrm{i} \Omega_m\eta_\mathrm{OM} \bar{N}_s [x_m,\rho_m]
\\
&+\gamma_m\Gamma_- \mathcal{D}_{a_m}[\rho_m]+\gamma_m\Gamma_+ \mathcal{D}_{a_m^\dagger}[\rho_m],\nonumber
\end{align}
where the heating and cooling rates $\gamma_m\Gamma_\pm= \Omega_m^2\eta_\mathrm{OM}^2\mathrm{Re}\{d_\mp|_{\gamma_m=0}\}$ coincide precisely with those defined in the main text. This master equation has a very simple Gaussian steady state $\bar{\rho}_m$ corresponding to a displaced thermal state with mean
\begin{equation}\label{displacement} 
\langle a_m\rangle = \frac{\mathrm{i} \Omega_m\eta_\mathrm{OM}\bar{N}_s}{\mathrm{i}\Omega_m+\gamma_m}\approx\eta_\mathrm{OM}\bar{N}_s,
\end{equation}
phonon number $\langle\delta a_m^\dagger \delta a_m\rangle=\bar{n}_m$, where $\bar{n}_m$ is given by Eq.~(4) in the main text, and $\langle\delta a_m^2\rangle=0$. We note that, starting from a thermal state, the mechanical mode relaxes to this steady state with a rate $\gamma_\mathrm{eff}=\gamma_m(1+\Gamma)$, where $\Gamma=\Gamma_--\Gamma_+$ is what we have called the cooling efficiency in the main text, since the equations of motion for the phonon number fluctuations and the mechanical field amplitude are given by
\begin{eqnarray}
&&\partial_t \Bra \delta a_m^\dagger \delta a_m\Ket=-2\gamma_m(1+\Gamma)\,\Bra \delta a_m^\dagger \delta a_m\Ket+2 \gamma_m(\bar{n}_\mathrm{th}+\Gamma_-),\nonumber
 \\
&&\partial_t \Bra a_m\Ket \hspace{-0.5mm}=\hspace{-0.5mm} [-\mathrm{i} \Omega_m \hspace{-0.5mm}-\hspace{-0.5mm}\gamma_m(1\hspace{-0.5mm}+\hspace{-0.5mm}\Gamma)]\,\Bra a_m\Ket\hspace{-0.5mm}+\hspace{-0.5mm}\mathrm{i}\eta_\mathrm{OM}\Omega_m \Bra a_s^\dagger a_s\Ket.
\end{eqnarray}
We have checked that this rate $\gamma_\mathrm{eff}$ is smaller than the decay rate of the optical correlator $s(\tau)$ for the parameters of interest, hence making Markov a valid approximation.
 
Let us remark that throughout the work we have been using both Eq.~(\ref{NZE2}) and Eq.~(\ref{ME Born-Markov plus RWA}) to obtain the steady-state moments of the mechanical oscillator. We have never observed any notable differences between them, except when working extremely close to threshold within the semi-classical approach, see
the inset of Fig.~3.  In these cases, however, the failure of Eq.~(\ref{ME Born-Markov plus RWA}) can be directly attributed to the failure of the semi-classical approach, and not to a failure of the rotating-wave approximation itself, which indeed is very well satisfied as shown by the c-MoP approach. Thus, we conclude that for the parameter regime studied in this work (compatible with current cWGM resonators) the state of the mechanical oscillator is indeed a displaced thermal state, with a phonon number that can only be evaluated once the optical photon number $\bar{N}_s$ and correlation function $s(\tau)$ are known.

\section{Semi-classical approach}\label{Sec Semi-Class SupMat}

The simplest way of obtaining the optical correlator $s(\tau)$ is by using standard linearization on $\mathcal{L}_\mathrm{opt}$. In this approach, we move to a displaced picture in which the large coherent background of the pump mode is removed, and then keep terms of the transformed optical Liouvillian only up to second order in the the bosonic operators. The displacement operator $D=\exp[\epsilon_p(a_p-a_p^\dagger)/\gamma_0]$ allows us to move to the new picture, in which the transformed optical state $\tilde{\rho}_\mathrm{opt}=D^\dagger\rho_\mathrm{opt}D$ evolves according to a transformed Liouvillian $\tilde{\mathcal{L}}_\mathrm{opt}=D^\dagger\mathcal{L}_\mathrm{opt}D$. Removing terms beyond quadratic order, this transformed Liouvillian can be written as a sum of independent Liouvillians for the pump and signal modes, $\tilde{\mathcal{L}}_\mathrm{opt}=\mathcal{L}_p+\mathcal{L}_s$, with $\mathcal{L}_p=\gamma_0\mathcal{D}_{a_p}$ and
\begin{equation}\label{signal Liouvillian App}
\gamma_0^{-1}\mathcal{L}_s(\cdot)=\left[-\mathrm{i}\Delta a_s^\dagger a_s+\frac{\sigma}{2}(a_s^{\dagger 2}-a_s^2),\cdot\right]+\mathcal{D}_{a_s}[\cdot], 
\end{equation}
with the injection parameter $\sigma=\epsilon_p \chi/\gamma_0^2$ and normalized detuning $\Delta=\Delta_s/\gamma_0$. Consequently, the optical steady state in the original picture becomes the separable state $\bar{\rho}_\mathrm{opt}=|\epsilon_p/\gamma_0\rangle\langle\epsilon_p/\gamma_0|\otimes\bar{\rho}_s$ where $|\epsilon_p/\gamma_0\rangle$ is a coherent state of amplitude $\epsilon_p/\gamma_0$ and $\bar{\rho}_s$ is the Gaussian state satisfying $\mathcal{L}_s[\bar{\rho}_s]=0$. The latter is completely characterized by its first and second moments, which are trivially found to be $\langle a_s\rangle=0$, $\langle a_s^\dagger a_s\rangle=\sigma^2/2(1+\Delta^2-\sigma^2)\equiv\bar{N}_s$, and $\langle a_s^2\rangle=\sigma(1-\mathrm{i}\Delta)/2(1+\Delta^2-\sigma^2)$, where we use the usual notation $\langle A\rangle=\mathrm{tr}_s\{A\rho_s\}$ for any operator $A$ acting on the signal subspace.

The optical correlation function simplifies to $s(\tau)=\text{tr} \{a_s^\dagger a_s e^{\mathcal L_s \,\tau} \mu_s\}$ where we have defined a traceless operator $\mu_s=(a_s^\dagger a_s-\bar{N}_s)\bar{\rho}_s$. Using again the fact that the Liouvillian $\mathcal{L}_s$ is Gaussian, it is simple to evaluate the correlation function $s(\tau)$. To this aim, let us define the column vector
\begin{equation}
\vec{v}(\tau)= \mathrm{col}\left(\widetilde{\Bra a_s^\dagger a_s\Ket},\widetilde{\Bra a_s^2\Ket},\widetilde{\Bra a_s^{\dagger\,2}\Ket}\right),
\end{equation}
where the expectation value of an operator $A$ with the tilde is defined as $\widetilde{\langle A\rangle}=\mathrm{tr}\{Ae^{\mathcal{L}_s\tau}\mu_s\}$. Taking the derivative of this vector with respect to $\tau$, we find the linear system $\partial _\tau \vec{v}(\tau)= L \vec{v}(\tau)$, where the matrix $M$ reads
\begin{equation}
  L= \gamma_0
  \left( {\begin{array}{ccc}
   -2  & \sigma & \sigma \\
   2 \sigma & -2 (1+i\Delta) & 0\\
   2 \sigma & 0 & -2 (1-i\Delta)\\
  \end{array} } \right).	
\end{equation}
It is straightforward to solve this linear system, for example by diagonalizing $L$. We write $L = U \Lambda U^{-1}$, with a similarity matrix $U$ that can be found analytically (but its expression is too lengthy to be reported here), and a diagonal matrix $\Lambda$ containing the eigenvalues of $L$, $\lambda_1=-2 \gamma_0$, and $\lambda_{2,3}= -2\gamma_0(1\pm \mathrm{i} \sqrt{\Delta^2-\sigma^2})$. Notice that for $\sigma>\Delta$ the square root becomes imaginary, making $\lambda_2<\gamma_0$, and in fact $\lambda_2=0$ at threshold, $\sigma=\sqrt{1+\Delta^2}$. Consequently, we call the region with $\sigma>\Delta$ the \textit{critical slowing down regime}. The solution of the linear system is then found as
\begin{equation}
\vec{v}(\tau)=U e^{\Lambda \tau} U^{-1} \vec{v}(0)\equiv \sum_{n=1}^3 L_n e^{\lambda_n \tau}\, \vec{u}, \label{v_sol}
\end{equation} 
where we have defined the initial condition vector
\begin{eqnarray}\label{signal cf vec}
 \vec{u}&&=\vec{v}(0)=\mathrm{col}\left(\langle a_s^\dagger a_s a_s^\dagger a_s \rangle-\bar{N}_s^2,\right.
\\
&&\hspace{1.5cm}\left.\langle a_s^2 a_s^\dagger a_s \rangle-\langle a_s^2 \rangle\bar{N}_s, \langle a_s^{\dagger 3} a_s \rangle-\langle a_s^2 \rangle^*\bar{N}_s\right),\nonumber
\end{eqnarray}
and the matrices $L_n = U \Pi_n U^{-1}$, where $(\Pi_n)_{jl}=\delta_{jn}\delta_{ln}$. Note that the vector $\vec{u}$ is formed by fourth order moments. In order to find them, we simply exploit the Gaussian structure of $\mathcal{L}_s$, which allows us to express moments of any order as products of moments of first and second order. Specifically, concerning third and fourth order moments we simply use
\begin{subequations}\label{GSAmoments}
\begin{eqnarray}
\langle \delta a_s^\dagger \delta a_s^2 \rangle &=& \langle \delta a_s^3 \rangle = 0
\\
\langle \delta a_s^{\dagger 2} \delta a_s^2 \rangle &=& \langle \delta a_s^{\dagger 2}\rangle\langle \delta a_s^2\rangle + 2 \langle \delta a_s^\dagger \delta a_s\rangle^2,
\\
\langle \delta a_s^\dagger \delta a_s^3\rangle &=& 3\langle\delta a_s^\dagger \delta a_s\rangle\langle \delta a_s^2\rangle,
\end{eqnarray}
\end{subequations}
where $\delta a_s=a_s-\langle a_s\rangle$. Note that the optical correlation function we are looking for is given by the first component of the vector, $s(\tau)=[\vec{v}(\tau)]_1$, and the integrals appearing in $d_0$ and $d_\pm$ in Eq.~(\ref{d1 and d2}) can be easily evaluated due to the exponential $\tau$-dependence of $\vec{v}(\tau)$ in Eq.~(\ref{v_sol}). 

\section{Cooling by heating via a two-photon process}\label{cooling by heating App}

In the main text we have shown that significant cooling can be obtained when working in the resolved sideband regime $\Omega=\Omega_m/\gamma_0\gg1$ and close to the resonance condition $\delta_\mathrm{eff}=\sqrt{\Delta^2-\sigma^2}=\Omega/2$. We have interpreted such a phenomenon as a ``cooling by heating'' effect, for which we have moved to a new picture defined by the squeezing operator $S(r)=\exp[-\mathrm{i}r(a_s^{\dagger 2} + a_s^2)/2]$ within the semi-classical approach explained above. Here we want to explicitly perform an adiabatic elimination of the optical mode in this picture, what will allow us to get more insight into the cooling mechanism.

Let us first write the master equation in this ``squeezed picture''. The transformation diagonalizes the Hamiltonian in the optical Liouvillian (\ref{signal Liouvillian App}), turning it into
\begin{eqnarray}\label{squeezing trafo}
&&\gamma_0^{-1}S^\dagger(r)\mathcal{L}_\mathrm{opt} S(r)[\cdot] = \mathrm{i} [\delta_{\text{eff}} a_s^\dagger a_s,\cdot]
\\
&&\hspace{3mm}+(1+\bar{N}_\mathrm{eff})\mathcal{D}_{a_s}[\cdot]+\bar{N}_\mathrm{eff} \mathcal{D}_{a_s^\dagger}[\cdot]+\mathrm{i}M \mathcal{K}_{a_s}[\cdot]-\mathrm{i}M \mathcal{K}_{a_s^\dagger}[\cdot],\nonumber
\end{eqnarray}
where we have defined the superoperator $\mathcal{K}_J[\cdot]=2 J(\cdot)J-J^2(\cdot)-(\cdot)J^2$, as well as the parameters $\bar{N}_\mathrm{eff}=(\Delta/\delta_{\text{eff}}-1)/2$ and $M=\sigma/2\delta_{\text{eff}}$. Note that the $\mathcal{K}$ terms rotate at frequency $2\delta_\mathrm{eff}$, and hence are highly suppressed when we work within the cooling condition $\delta_\mathrm{eff}\approx \Omega/2$ and with $M/2\delta_\text{eff}=\sigma/4\delta_\text{eff}^2\ll 1$ (rotating-wave approximation). Therefore, as mentioned in the main text, in this picture the signal field is turned into a bosonic mode with oscillation frequency $\delta_\mathrm{eff}$, and at thermal equilibrium with occupation $\bar{N}_\mathrm{eff}$. On the other hand, the photon-number operator is transformed into $S^\dagger(r)a_s^\dagger a_s S(r)=\bar{N}_\mathrm{eff} + (2 \bar{N}_\mathrm{eff}+1) a_s^\dagger a_s+\mathrm{i} M (a_s^2-a_s^{\dagger \,2})$, and hence the optomechanical interaction can be approximated by
\begin{equation}\label{transOMcoupling}
S^\dagger(r)a_s^\dagger a_s S(r)(a_m+a_m^\dagger) \approx \mathrm{i} M(a_s^2 a_m^\dagger -a_s^{\dagger 2} a_m).
\end{equation} 
within the rotating-wave approximation as long as $\eta_\text{OM}(2\bar{N}_\text{eff}+1)=\eta_\text{OM}\Delta/\delta_\text{eff}\ll 1$.

Hence, within these conditions, the transformed state $\tilde{\rho}=S^\dagger(r)\rho S(r)$ evolves according to a master equation that we write as
\begin{equation}
\partial_t \tilde{\rho} = \tilde{\mathcal{L}}_\mathrm{s}[\tilde{\rho}]+\mathcal{L}_m[\tilde{\rho}]+\tilde{\mathcal{L}}_\mathrm{OM}[\tilde{\rho}]
\end{equation}
with
\begin{subequations}
\begin{eqnarray}
\tilde{\mathcal{L}}_\mathrm{s}[\tilde{\rho}] &=& \left[-\mathrm{i}\gamma_0\delta_\mathrm{eff} a_s^\dagger a_s,\tilde{\rho}\right]
\\
&&+\gamma_0(\bar{N}_\mathrm{eff}+1) \mathcal{D}_{a_s}[\tilde{\rho}]+\gamma_0\bar{N}_\mathrm{eff}\mathcal{D}_{a_s^\dagger}[\tilde{\rho}],\nonumber
\\
\mathcal{L}_m[\tilde{\rho}] &=&[-\mathrm{i}\Omega_m a_m^\dagger a_m,\tilde{\rho}]
\\
&&+\gamma_m (\bar{n}_\mathrm{th}+1)\mathcal{D}_{a_m}[\tilde{\rho}]+\gamma_m \bar{n}_\mathrm{th}\mathcal{D}_{a_m^\dagger}[\tilde{\rho}],\nonumber
\\
\tilde{\mathcal{L}}_\mathrm{OM}[\tilde{\rho}] &=& [\Omega_m\eta_\mathrm{OM}M(a_s^2 a_m^\dagger-a_s^{\dagger 2} a_m),\tilde{\rho}],
\end{eqnarray}
\end{subequations}
where $\bar{N}_\mathrm{eff}$ and $M$ are defined in the main text. The structure of this master equation is similar to the original one, Eq. (\ref{App full master equation}), with the only difference that the optical Liouvillian is replaced by $\tilde{\mathcal{L}}_s$, corresponding to a single-mode at finite temperature, and the optomechanical interaction $a_s^\dagger a_s (a_m+a_m^\dagger)$ by $\mathrm{i}M(a_s^2 a_m^\dagger-a_s^{\dagger 2} a_m)$. The adiabatic elimination of the optical mode can be carried out exactly in the same way as we did in Appendix \ref{App1}, and under the cooling condition $\delta_\mathrm{eff}=\Omega/2$ it would lead to the heating and cooling rates
\begin{subequations}
\begin{eqnarray}
\Gamma_-&\approx& \frac{1}{2}CM^2\mathrm{tr}\{a_s^2 a_s^{\dagger 2}\tilde{\rho}_s\},
\\
\Gamma_+&\approx& \frac{1}{2}CM^2\mathrm{tr}\{a_s^{\dagger 2} a_s^2\tilde{\rho}_s\},
\end{eqnarray}
\end{subequations}
where $C=\Omega_m^2\eta_\mathrm{OM}^2/\gamma_m\gamma_0$ is the bare optomechanical cooperativity, and $\tilde{\rho}_s$ is a thermal state with mean photon number $\bar{N}_\mathrm{eff}$. The cooling efficiency is then given by
\begin{equation}
\Gamma=\Gamma_--\Gamma_+=\frac{1}{2}CM^2\mathrm{tr}\{[a_s^2,a_s^{\dagger 2}]\tilde{\rho}_s\}=2CM^2\hspace{-1mm}\left(\hspace{-0.5mm}\bar{N}_\mathrm{eff}\hspace{-0.5mm}+\hspace{-0.5mm}\frac{1}{2}\hspace{-0.5mm}\right)\hspace{-1mm}.
\end{equation}
The ``cooling by heating'' effect is clearly seen because the cooling efficiency increases with the effective thermal photon number $\bar{N}_\mathrm{eff}$. But it is important to note that this enhancement of the cooling efficiency is a direct consequence of the commutator appearing in the trace, contributing as $[a_s^2,a_s^{\dagger 2}]=4a_s^\dagger a_s+2$, which in turn comes from the fact that the effective optomechanical interaction $\mathrm{i}(a_s^2 a_m^\dagger-a_s^{\dagger 2} a_m)$ corresponds to the exchange of phonons with pairs of photons. In the usual sideband laser cooling scenario, the effective optomechanical interaction is bilinear, e.g. $\mathrm{i}(a_s a_m^\dagger-a_s^\dagger a_m)$, meaning that the commutator in the expression above is replaced by $[a_s,a_s^\dagger]=1$, and hence the thermal photonic background does not enter the cooling efficiency.

Let us finally note that the fundamental limit can be written as
\begin{equation}
\bar{n}_\mathrm{FL}=\frac{\Gamma_+}{\Gamma}=\frac{\mathrm{tr}\{a_s^{\dagger 2} a_s^2\tilde{\rho}_s\}}{\mathrm{tr}\{[a_s^2,a_s^{\dagger 2}]\tilde{\rho}_s\}}=\frac{\bar{N}_\mathrm{eff}^2}{2\bar{N}_\mathrm{eff}+1}\;\xrightarrow[]{\bar{N}_\mathrm{eff}\gg1} \; \frac{\bar{N}_\mathrm{eff}}{2},
\end{equation}
which increases linearly with the effective thermal photon number. Hence, as explained in the main text, the ``cooling by heating'' mechanism is optimized by finding a proper trade off between the increase in the cooling efficiency (``good noise'') and the increase in the fundamental limit (``bad noise''). It is to be noted that within the usual sideband laser cooling, any thermal background will still contribute to this fundamental limit, $\bar{n}_\mathrm{FL}=\mathrm{tr}\{a_s^\dagger a_s \rho_s\}/\mathrm{tr}\{[a_s,a_s^\dagger]\rho_s\}\xrightarrow[]{\bar{N}_\mathrm{eff}\gg1}\bar{N}_\mathrm{eff}$, but, as explained above, it provides no enhancement of the cooling efficiency $\Gamma$. In other words, in standard sideband cooling the thermal background acts only as ``bad'' noise.

We emphasize that these expressions for $\Gamma$ and $\bar{n}_\mathrm{FL}$ agree with the ones provided in the main text, which have been first calculated exactly within the semi-classical approach, and then approximated to leading order in $1/\Omega^2$ for $\delta_\mathrm{eff}=\Omega/2$. 

\begin{figure*}
\includegraphics[width=\textwidth]{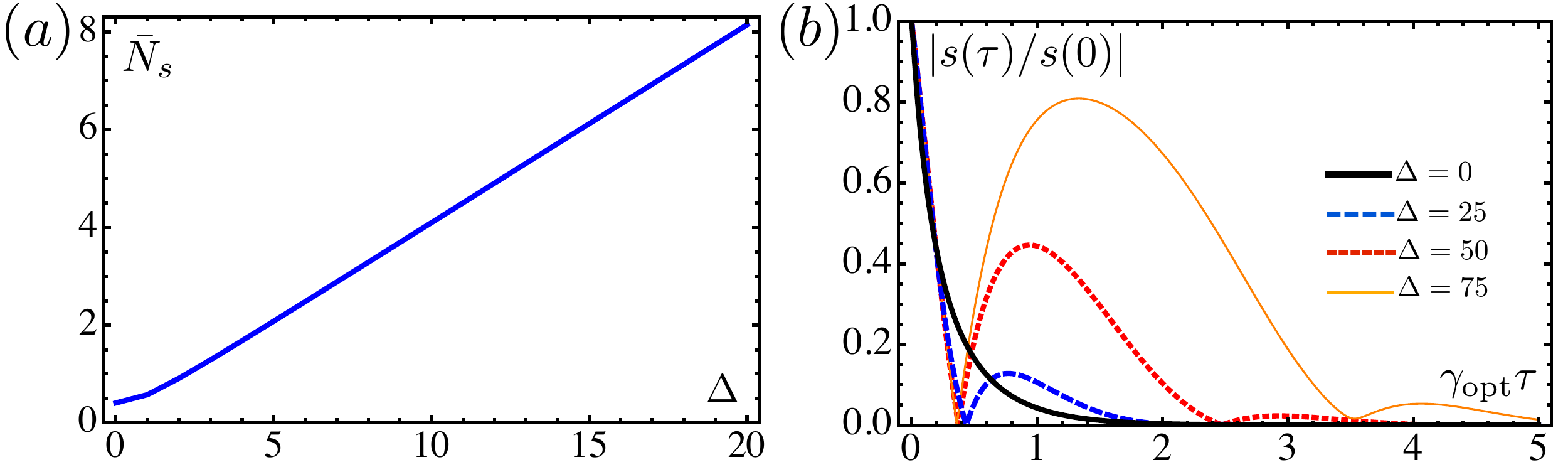}
\caption{\label{scaling} (Color online) (a) Steady state photon number $\bar{N}_s=\mathrm{tr}\{a_s^\dagger a_s\bar{\rho}_\mathrm{opt}\}$ at the critical point as a function of the normalized detuning $\Delta$. (b) Absolute value of the normalized optical correlation function $s(\tau)$ at the critical point for different values of detuning as of function of the normalized time $\gamma_\mathrm{opt}\tau$, with $\gamma_\mathrm{opt}=\gamma_0\eta_\mathrm{DC}(1+\Delta)$. In both figures the parameters are $\gamma_0=1$ and $\eta_\mathrm{DC}=0.01$, and we have obtained them by applying Gaussian c-MoP theory to the optical problem.}
\end{figure*}

\section{c-MoP approach to the optical problem \\ and scalings at the critical point}\label{c-MoP theory}

Despite the analytical and physical insight that it provides, the semi-classical approach suffers from several issues, most importantly its divergent character at threshold, which shows that it cannot be trusted when working close to such point. Unfortunately, there is no systematic way of checking within the formalism itself where exactly it fails. This question can only be answered by comparing it to a more accurate approach. To this aim, we have applied our recently developed self-consistent Mori projector (c-MoP) theory \cite{Degenfeld14,Degenfeld15}. In particular this approach has allowed us to characterize optical steady-state observables such as the photon number $\bar{N}_s$ or the correlation function $s(\tau)$ in all relevant parameter space, including threshold.

We detailed the application of c-MoP theory \cite{Degenfeld14} to the DOPO problem in \cite{Degenfeld15}, including its combination with the Gaussian-state approximation that we use in this work, which was shown to be quite accurate both for steady-state quantities and dynamics. Let us now briefly introduce such an approach here for completeness, keeping in mind that details can be looked up in \cite{Degenfeld15}. Our main goal consists in finding the optical correlation function (\ref{two time corr}), for which we need to solve the dynamics generated by the optical Liouvillian $\mathcal{L}_\mathrm{opt}$, that is, the DOPO dynamics
\begin{equation}\label{masterDOPO}
\dot{\rho}_\mathrm{opt}=\mathcal{L}_\mathrm{opt}\rho_\mathrm{opt}.
\end{equation}
This is a nonlinear two-mode problem with no analytic solution and whose direct numerical simulation becomes unfeasible for moderate photon numbers already. In contrast, c-MoP theory works with the reduced states of the pump and signal modes, $\rho_p(t)=\mathrm{tr}_s\{\rho_\mathrm{opt}\}$ and $\rho_s(t)=\mathrm{tr}_p\{\rho_\mathrm{opt}\}$, respectively, whose coupled dynamics are approximated by the set of nonlinearly coupled equations
\begin{subequations}\label{cMoP eq app}
\begin{eqnarray}
\dot{\rho}_s(t)&=&\mathcal{L}_s\rho_s(t)+\frac{\chi}{2}\left[a_s^{\dagger 2}\Bra a_p\Ket(t)-a_s^2\Bra a_p^\dagger\Ket(t),\rho_s(t)\right]\nonumber 
\\
&&+\left( \frac{\chi}{2}\right)^2\left\{\left[a_s^2\,,\,h_s(t)\right]+\text{H.c.}\right\},
\\
\dot{\rho}_p(t)&=&\mathcal{L}_p \rho_p(t)+\frac{\chi}{2}\left[a_p \Bra a_s^{\dagger\,2}\Ket(t)-a_p^\dagger\Bra a_s^2\Ket(t)\,,\rho_p(t)\right]\nonumber 
\\
&&+\left(\frac{\chi}{2}\right)^2\left\{\left[a_p\,,\sum_{n=1}^3 h_{p,n}(t)\right]+\text{H.c.}\right\},
\\
\dot h_s(t)&=& (-\gamma_p +\mathcal L_s) h_s(t)  + \mathcal K _s(t,t)\rho_s(t),
\\
\dot h_{p,n}(t)&=& (\lambda_n +\mathcal L_p) h_{p,n}(t)+\mathcal K_{p,n}(t,t)\rho_p(t),
\end{eqnarray}
\end{subequations}
where $h_s(t)$ and $\{h_{p,n}(t)\}_{n=1,2,3}$ are auxilliary operators acting on the signal and pump subspaces (introduced to turn the c-MoP equations into ordinary differential equations, since originally they have an integro-differential structure \cite{Degenfeld15}), and we refer to \cite{Degenfeld15} for the definitions of the superoperators $\mathcal{L}_s$, $\mathcal{L}_p$, $\mathcal{K}_s(t,t')$, and $\mathcal{K}_{p,n}(t,t')$. Denoting by $D_s$ and $D_p$ the dimensions of the truncated Hilbert spaces of signal and pump in a numerical simulation, we see that the original problem (\ref{masterDOPO}) requires solving a $D_s\times D_p$ system, whereas there are only $2D_s+4D_p$ c-MoP equations (or even less by exploiting the quadratic or Gaussian structure of the pump equations).

Nevertheless, even though c-MoP allows to get numerical insight in a larger region of parameter space, the simulation of problems with very large photon numbers (such as the ones we work with close to threshold) is still challenging. It is in those regions where a Gaussian-state approximation becomes extremely useful. As the name suggests, such approximation consists in assuming that the reduced signal and pump states are Gaussian, meaning that they are completely characterized by first and second order moments. Under such circumstances, we can approximate third and fourth order moments of $\bar{\rho}_s$ as in (\ref{GSAmoments}), and the c-MoP equations provide a closed set of nonlinear equations for the first and second order moments of the operators $\rho_s(t)$, $\rho_p(t)$, $h_s(t)$ and $\{h_{p,n}(t)\}_{n=1,2,3}$. The steady-state moments can then be efficiently found simply by finding the stationary solutions of these equations.

As an example, in Fig. \ref{scaling}a we show the steady-state photon number $\bar{N}_s$ at the critical point $(x=1)$ as a function of the normalized detuning $\Delta$. It shows a clear linear dependence for $\Delta>1$ which, together with the well-known $\eta_\mathrm{DC}^{-1}$ scaling with the down-conversion coupling \cite{Wolinsky88,Kinsler95,Veits97,Degenfeld15}, provides an overall $\bar{N}_s\propto(1+\Delta)/\eta_\mathrm{DC}$ scaling of the signal photon number at threshold. The knowledge of this scaling plays an important role for the determination of the conditions under which mechanical backaction on the optics can be neglected, as we have explained in Sec. \ref{BackAction}.

Let us now explain how the optical correlation function can be evaluated within this framework. First, note that we can rewrite it as
\begin{equation}
s(\tau)=\bar{N}_s \,(\mathrm{tr}\{a_s^\dagger a_s\nu(\tau)\}-\bar{N}_s)
\end{equation}
where $\nu(t)=e^{\mathcal{L}_\mathrm{opt}t}\nu(0)$ can be interpreted as an operator with evolution equation $\dot{\nu}=\mathcal L_{\text{opt}}\nu$ and initial condition $\nu(0)=a_s^\dagger a_s\bar{\rho}_\text{opt}/\bar{N}_s$. Since this evolution equation is formally equivalent to the optical master equation (\ref{masterDOPO}), we can apply c-MoP theory directly on $\nu(t)$, approximating it by a separable operator $\nu_s(t)\otimes\nu_p(t)$, with $\nu_p(t)=\mathrm{tr}_s\{\nu(t)\}$ and $\nu_s(t)=\mathrm{tr}_p\{\nu(t)\}$ evolving according to equations (\ref{cMoP eq app}) with $\rho_j$ replaced by $\nu_j$. Under a Gaussian approximation for $\nu_s(t)$ similar to (\ref{GSAmoments}) but with expectation values defined with respect to $\nu_s(t)$, the evolution equations for the first and second moments of $\nu_s(t)$, $\nu_p(t)$, $h_s(t)$, $\{h_{p,n}(t)\}_{n=1,2,3}$, and their Hermitian conjugates (note that $\nu$ is not Hermitian) form a closed nonlinear system which we can solve again efficiently. Note that the initial conditions for these moments, e.g., $\mathrm{tr}\{a_s^\dagger a_s\nu(0)\}=\mathrm{tr}\{a_s^\dagger a_s a_s^\dagger a_s\bar{\rho}_\text{opt}\}/\bar{N}_s$, are found from the Gaussian c-MoP steady-state solutions as explained above.

In Fig. \ref{scaling}b we show the evolution of the absolute value of the correlation function $s(\tau)$ at the critical point and for different values of the normalized detuning $\Delta$. Time is normalized to $[\gamma_0\eta_\mathrm{DC}(1+\Delta)]^{-1}$, and hence the fact that all the curves decay on the same time scale proves that the optical relaxation time scales as $\gamma_\mathrm{opt}=\gamma_0\eta_\mathrm{DC}(1+\Delta)$ at threshold. This again plays a fundamental role when proving that mechanical backaction is negligible, as we discussed in Sec. \ref{BackAction}.

\end{document}